\documentclass[prl,twocolumn,amsmath,nobibnotes,nofootinbib,superscriptaddress,preprintnumbers,hyperref]{revtex4-1}

\usepackage{bm} \usepackage{graphicx} \usepackage{color}
\usepackage{epstopdf} \usepackage{amssymb}
\def\ba{\begin{eqnarray}}
\def\ea{\end{eqnarray}}
\def\be{\begin{equation}}
\def\ee{\end{equation}}
\def\({\left(}
\def\){\right)}
\def\[{\left[}
\def\]{\right]}
\def\<{\left<}
\def\>{\right>}

\begin{document}

\title{A robust constraint on cosmic textures from the cosmic microwave background} 
\date{\today}Ê

\author{Stephen M. Feeney}
\email{stephen.feeney.09@ucl.ac.uk}
\affiliation{Department of Physics and Astronomy, University College London, London WC1E 6BT, U.K.}
\author{Matthew C. Johnson}
\email{mjohnson@perimeterinstitute.ca}
\affiliation{Perimeter Institute for Theoretical Physics, Waterloo, Ontario N2L 2Y5, Canada}
\author{Daniel J. Mortlock}
\email{d.mortlock@imperial.ac.uk}
\affiliation{Astrophysics Group, Imperial College London, Blackett Laboratory, Prince Consort Road, London, SW7 2AZ, U.K.}
\affiliation{Department of Mathematics, Imperial College London, London, SW7 2AZ, U.K.}
\author{Hiranya V. Peiris}
\email{h.peiris@ucl.ac.uk}
\affiliation{Department of Physics and Astronomy, University College London, London WC1E 6BT, U.K.}

\begin{abstract}
Fluctuations in the cosmic microwave background (CMB) contain information which has been pivotal in establishing the current cosmological model. These data can also be used to test well-motivated additions to this model, such as cosmic textures. Textures are a type of topological defect that can be produced during a cosmological phase transition in the early universe, and which leave characteristic hot and cold spots in the CMB. We apply Bayesian methods to carry out a rigorous test of the texture hypothesis, using full-sky data from the Wilkinson Microwave Anisotropy Probe. We conclude that current data do not warrant augmenting the standard cosmological model with textures. We rule out at 95\% confidence models that predict more than 6 detectable cosmic textures on the full sky.
\end{abstract}

\preprint{}

\maketitle

{\bf Introduction.} Precision measurements of anisotropies in the cosmic microwave background (CMB) radiation have been instrumental in establishing the standard ``$\Lambda$CDM'' model of cosmology: that the universe is composed mostly of dark energy and dark matter, with structures seeded by nearly scale-invariant Gaussian density fluctuations. In addition to establishing $\Lambda$CDM, the CMB is also an ideal observable for determining if there are departures from this baseline model.

In this paper, we present a novel algorithm for the analysis of CMB data from the Wilkinson Microwave Anisotropy Probe (WMAP)~\cite{Bennett:2003ba} to search for the presence of a class of topological defects known as cosmic textures~\cite{Turok:1989ai}. Although textures (and other topological defects, such as cosmic strings) have been ruled out as the dominant source for the primordial perturbations~\cite{Albrecht:1997mz,Albrecht:1997nt}, their production is inevitable in theories in which a non-Abelian global symmetry is broken~\cite{Kibble:1976sj}. Previous work~\cite{Cruz:2007pe,2008MNRAS.390..913C,Cruz:2009nd,2011MNRAS.410...33V} presented evidence, based on the properties of a single feature in the CMB, that $\Lambda$CDM should be augmented by adding cosmic textures. Implementing Bayesian model selection using data on the full sky, we are able to put the texture hypothesis to a much more stringent test. Incorporating this extra information, we conclude that the WMAP 7-year data do not warrant augmenting $\Lambda$CDM with cosmic textures, and place constraints on theories giving rise to textures. Our algorithm is easily extendable to incorporate better data, multiple datasets, and a more complete theoretical understanding of the properties and evolution of cosmic textures.

{\bf Cosmic texture theory.} The theory of cosmic textures posits a phase transition in the early universe in which a non-Abelian global symmetry is broken. In an expanding universe, different regions of the universe can be out of causal contact, obstructing the symmetry-breaking phase transition from occurring in the same manner everywhere in space~\cite{Kibble:1976sj}. Therefore, a scale-invariant set of knots in the symmetry-breaking order parameter inevitably form: these are cosmic textures. Subsequent to the phase transition, knots from the distribution come into causal contact with their surroundings and undergo collapse~\cite{Turok:1989ai,Turok:1991qq,Spergel:1990ee,Pen:1993nx,Turok:1990gw}. Upon collapse, textures unwind when the gradient energy of the field configuration exceeds the energy required to restore the global symmetry. As the field re-orders, the energy of the texture configuration is released as an outgoing shell of scalar field radiation. 

The gravitational potential associated with a cosmic texture varies in time as it collapses and subsequently explodes. CMB photons passing through an evolving texture will be redshifted if they pass through a collapsing texture, and blueshifted if they pass through an exploding texture~\cite{Turok:1990gw}. Each texture unwinding event therefore produces an additive hot or cold spot on the sky, which can be approximated as a disc whose angular size, $\theta_c$, depends on the distance to the texture unwinding event and whose amplitude, $\epsilon \equiv 8 \pi^2 G \eta^2$, depends on the scale of symmetry breaking $\eta$. The temperature profile in the central region of an unwinding event situated at the Galactic North Pole can be approximated as~\cite{Turok:1990gw}
\begin{equation}
t(\theta, \phi) =  \frac{(-1)^{p} \, \epsilon}{\sqrt{1+4 \left( \frac{\theta}{\theta_c} \right)^2 }},
\label{eq-texture-template}
\end{equation}
where $\theta$ and $\phi$ are Galactic co-latitude and longitude, respectively, and $p=\{0,1\}$. Here $p = 0$ corresponds to a hot spot and $p = 1$ to a cold spot. The form of the modulation for large $\theta$ is presently unknown; following Refs.~\cite{Cruz:2007pe,2008MNRAS.390..913C}, we match onto a Gaussian profile at the half-maximum radius, $\theta_* = \sqrt{3/2}\theta_c$.

The angular scale distribution is determined by the evolution of the cosmological horizon during the matter-dominated era~\cite{Turok:1990gw}, which is fixed by the late-time cosmological parameters of the $\Lambda$CDM model. In addition, each feature is equally likely to be: (i) hot or cold, and (ii) located at any point on the sky, allowing us to define the prior over the ``local'' template parameters as
\begin{equation}
{\rm Pr}(p,\theta_c,\theta_0, \phi_0 | \epsilon) = \frac{\sin \theta_0}{4 \pi \theta_c^3} \left( \frac{1}{{(2^\circ)}^2}  - \frac{1}{{(50^\circ)}^2} \right)^{-1} ,
\label{eq-local-priors}
\end{equation}
where $0 \le \theta_0 \le \pi$, $0 \le\phi_0 < 2\pi$, and we take $2^{\circ} \leq \theta_c \leq 50^{\circ}$. The lower limit on $\theta_c$ results from the large power on degree scales in the CMB; the upper limit stems from the fact that templates with $\theta_c > 50^\circ$ are large enough to cover the whole sky and overlap themselves, rendering Eq.~\ref{eq-texture-template} invalid.

Different theories giving rise to textures yield different predictions for the symmetry breaking scale and frequency of texture unwinding events; however, all mechanisms produce CMB modulations of the form described in Eq.~\ref{eq-texture-template}. Observationally, theories giving rise to textures are therefore differentiated only by the expected number of {\em detectable} texture unwinding events on the CMB sky, $\bar{N}_s$, and their amplitude, $\epsilon$. In our analysis, the background CMB fluctuations dominate the definition of detectability. The prior probability ${\rm Pr}(\bar{N}_s, \epsilon)$ is set by using simulations to determine the parameter space to which our algorithm is sensitive, as we will discuss shortly.

The $\Lambda$CDM+texture model can therefore be fully described by: the standard $\Lambda$CDM parameters; a set of ``global'' texture parameters, $\mathbf{m}_0 = \{ \bar{N}_s, \epsilon \}$, labelling theories; a set of ``local'' parameters, $\mathbf{m}_i = \{ p, \theta_c, \theta_0, \phi_0 \}_i$, describing each texture; and theoretical priors on these parameters, ${\rm Pr}(\mathbf{m}_0)$ and ${\rm Pr} (\mathbf{m}_i | \mathbf{m}_0)$. To test the $\Lambda$CDM+textures model against vanilla $\Lambda$CDM, we need only vary those parameters that are unique to the more complex model~\cite{dickey}. We therefore fix the $\Lambda$CDM parameters to their best-fit values from the analysis of WMAP 7-year data~\cite{Komatsu:2010fb} (hereafter referred to as WMAP7). We will now describe the specifics of our search algorithm.

{\bf Searching for textures.} The fundamental question posed by this analysis is: are the WMAP7 data better described by the standard $\Lambda$CDM cosmological model or $\Lambda$CDM plus cosmic textures? The goal is to calculate the joint posterior distribution of $\bar{N}_s$ and $\epsilon$, given the available data. Pure $\Lambda$CDM corresponds to $\bar{N}_s = 0$. We avoid the {\em a posteriori} selection effects associated with postdicting an explanation for anomalous portions of the data (see Ref.~\cite{Bennett:2010jb} for an in-depth discussion) by performing an analysis of the full dataset. This is important, given that previous evidence~\cite{2008MNRAS.390..913C} for cosmic textures in the CMB was based on the analysis of a single anomalous feature, the so-called CMB Cold Spot~\cite{Cruz:2009nd}.

Given an expected number of detectable textures over the whole sky, $\bar{N}_s$, the actual number of detectable textures, $N_s$, is drawn from a Poisson distribution with mean $f_{\rm sky} \bar{N}_s$, where $f_{\rm sky}$ is the fraction of the sky covered by the observations. The full posterior probability distribution of the global parameters describing the texture model, $\epsilon$ and $\bar{N}_s$, is given by marginalizing the likelihood, ${\rm Pr} (\mathbf{d} | \mathbf{m}_1, \ldots \mathbf{m}_{N_s}, \epsilon, N_s, f_{\rm sky} )$, weighted by the prior, over the (unknown) actual number of textures and their individual properties.  This is an extremely challenging integral to evaluate directly, but a good approximation to it can be found by identifying the regions of this parameter space in which the likelihood is appreciable and only including these contributions~\cite{Feeney:2010dd,Feeney:2010jj}. Extending this formalism to also incorporate the global parameter $\epsilon$ allows us to self-consistently combine the evidence that each candidate is a texture into a global constraint on the texture theory. The resultant expression (cf.\ Refs.~\cite{Feeney:2010dd,Feeney:2010jj}) is
\begin{widetext}
\begin{equation}
{\rm Pr}(\epsilon, \bar{N}_s | \mathbf{d}, f_{\rm sky} ) \simeq  \frac{{\rm Pr} (\epsilon, \bar{N}_s) {\rm Pr} (\mathbf{d} | N_s = 0, f_{\rm sky} ) }{ {\rm Pr} (\mathbf{d} | f_{\rm sky}) } e^{-f_{\rm sky} \bar{N}_s} 
\!
\sum_{N_s = 0}^{N_b} \frac{(f_{\rm sky} \bar{N}_s)^{N_s}}{N_s !} 
\!\!\!
\sum_{b_1, b_2, \ldots, b_{N_s} = 1}^{N_b}
\!\!
\Delta^{b_1 b_2 \ldots b_{N_s}} \prod_{s = 1}^{N_s} \rho_{b_s} (\epsilon) \ , 
\label{eq-approx-posterior}
\end{equation}
where ${\rm Pr} (\epsilon, \bar{N}_s)$ is the prior (the properties of which are discussed below), ${\rm Pr} (\mathbf{d} | N_s = 0, f_{\rm sky} )$ is the likelihood for $\Lambda$CDM (i.e.\ the likelihood assuming no textures), and $N_b$ denotes the number of  regions on the sky, or ``blobs," containing candidate signatures, each labeled by $b_i$. The actual number of detectable textures $N_s$ lies between $0$ and $N_b$. 
The quantity $\Delta^{b_1 b_2 \ldots b_{N_s}}$ is one when all indices take distinct values and zero otherwise: it generates all permutations of $N_s$ textures located in $N_b$ blobs, assuming no more than one texture per blob. Finally, the quantity $\rho_{b_i} (\epsilon)$, defined as
\begin{equation}
\rho_{b_i} (\epsilon) \equiv \sum_{p=0,1} \frac{\int_{b_i} {\rm d} \theta_0 {\rm d} \phi_0 \int {\rm d} \theta_c {\rm Pr}(p,\theta_c,\theta_0, \phi_0 | \epsilon) {\rm Pr} (\mathbf{d}_{b_i} | p,\theta_c, \theta_0,\phi_0, \epsilon, N_s=1, f_{\rm sky}) }{ f_{\rm sky} \, {\rm Pr} (\mathbf{d}_{b_i} | N_s = 0, f_{\rm sky} )},
\label{eq-evidence-ratio}
\end{equation}
\end{widetext}
is a patch-based evidence ratio evaluated in each blob: this is a measure of how much better $\Lambda$CDM plus a single texture fits the data than pure $\Lambda$CDM, considering only the data in blob $b_i$. The factor of $f_{\rm sky}$ appearing in the denominator accounts for the fact that we are restricted to detecting textures outside the sky cut. Unless the data provide strong support for the presence of a texture, the evidence ratio penalizes this more complicated model through the larger volume of parameter space that must be considered in constructing the priors, thus self-consistently implementing Occam's razor.
 
 The likelihood for blob $b_i$ is
\begin{eqnarray}
 && {\rm Pr} (\mathbf{d}_{b_i} | p,\theta_c, \theta_0,\phi_0, \epsilon, N_s=1, f_{\rm sky}) = \\  
&& \frac{1}{(2 \pi)^{N_{{\rm pix}, b_i} / 2} |\mathbf{C}_{b_i}|}
    e^{- [\mathbf{d}_{b_i} - \mathbf{t}(\epsilon, \mathbf{m}_1)] 
      \mathbf{C}_{b_i}^{-1} 
    [\mathbf{d}_{b_i} - \mathbf{t}(\epsilon, \mathbf{m}_1)]^{\rm{T}} / 2} ,
\nonumber
\label{eq-likelihood}
\end{eqnarray}
where $N_{{\rm pix}, b_i}$ is the total number of pixels in the blob, $\mathbf{d}_{b_i}$ are the data points in the blob, and $\mathbf{C}_{b_i}$ is the pixel-pixel covariance matrix using only pixels contained in the blob, which includes the fluctuations due to $\Lambda$CDM as well as instrumental noise and the effects of the beam. 

{\bf Locating texture candidates.} To evaluate Eq.~\ref{eq-approx-posterior}, we must first identify the most promising candidates in the map. We do so by employing the suite of spherical needlet transforms~\cite{Marinucci:2007aj,Pietrobon:2008rf,Scodeller:2010mp} defined in Ref.~\cite{Feeney:2010dd}. Filtering CMB temperature maps with spherical needlets yields information about both the position and angular size of interesting features. The statistics of the filtered field (established using 3000 simulated Gaussian CMB realizations) can then be used to assess the significance of a candidate. Applying the needlet transform to texture templates (Eq.~\ref{eq-texture-template}) of various sizes yields a lookup table specifying the needlet whose response is maximal at each texture size. This table can then be used to identify peaks in a filtered input map with a texture candidate of a certain size. To minimize the number of false detections, while not discarding potentially interesting signals, we determine a set of size-dependent thresholds (identical to those in Ref.~\cite{Feeney:2010dd}) using an end-to-end simulation of the WMAP experiment (see Refs.~\cite{Jarosik:2010iu,Gold:2010fm}) containing a $\Lambda$CDM CMB as well as realistic foregrounds and systematics that we cannot include in our likelihood function. The thresholds chosen restrict the number of candidate textures -- by definition false detections -- to be of order ten.  All thresholds and parameters in the needlet transform are fixed at this point.

{\bf Sensitivity testing.} To determine our ability to detect textures given our thresholds, we generate a set of CMB maps from the WMAP7 best-fit power spectrum, and place textures of varying $\epsilon$ and $\theta_c$ in both a region with low and high instrumental noise (as the noise properties of the WMAP experiment vary according to position on the sky). We find that, for $2^{\circ} \leq \theta_c \leq 50^{\circ}$, the significance threshold is certainly exceeded (and therefore a candidate identified) for $\epsilon > 10^{-4}$. For a favorable realization of the background CMB and instrumental noise, candidates are detected for $\epsilon > 2.5 \times 10^{-5}$ at scales $\theta_c \agt 5^{\circ}$ and $\epsilon > 5 \times 10^{-5}$ at somewhat smaller scales. We use $\epsilon = 2.5 \times 10^{-5}$ as a lower limit for detectable textures, and neglect the effect of $\theta_c$ on our candidate detection efficiency as it is far less important than the factor of $\theta_c^{-3}$ in Eq.~\ref{eq-local-priors}.

{\bf Calculating the texture posterior probability.} Once the candidate textures have been identified, the posterior probability distribution Eq.~\ref{eq-approx-posterior} can be calculated by first evaluating the patch-based evidence ratio Eq.~\ref{eq-evidence-ratio} for each blob using the MultiNest~\cite{Feroz:2008xx,Feroz:2007kg} nested sampling software. This requires calculating the inverse covariance matrix $\mathbf{C}_{b_i}^{-1}$, which is extremely memory-intensive at full WMAP resolution: the necessary storage capacity scales with size as $\theta_c^4$. We therefore employ an adaptive-resolution analysis, processing each blob at the highest resolution possible given its size and the available computational resources. This removes the limitation on blob size of Refs.~\cite{Feeney:2010dd,Feeney:2010jj}.

The only remaining quantity to evaluate in Eq.~\ref{eq-approx-posterior} is the prior ${\rm Pr} (\epsilon, \bar{N}_s)$. We choose a uniform prior for $\epsilon$ between $2.5 \times10^{-5} \leq \epsilon \leq 1.0 \times 10^{-4}$. The lower bound is an estimate of what is detectable with our pipeline, determined by the simulations described above. The upper bound comes from requiring that the symmetry-breaking scale for textures, $\eta$, is below the scale of cosmological inflation. To be consistent with the lack of observed B-mode polarization in the CMB~\cite{Larson:2010gs,Komatsu:2010fb}, the scale of inflation must be less than approximately $10^{16}$ GeV, constraining $\epsilon$ to be less than roughly $10^{-4}$ (this agrees with the prior of Ref.~\cite{Cruz:2007pe}).

We adopt a uniform prior on $\bar{N}_s$ between $0 \leq \bar{N}_s \leq 10$. The comoving density of textures produced in a phase transition depends on the particular texture model in question, and can be determined from simulations. The total number of unwinding events is obtained by integrating this density over the four-volume swept out by the CMB photons. For example, simulations~\cite{Cruz:2007pe} of SU(2) textures indicate that we can expect to have causal access to roughly 7 textures with $\theta_c > 2^{\circ}$ in the CMB. The number of these unwinding events which are then {\em detectable} is mainly a function of our particular realization of the background CMB. Our choice of a uniform prior accounts for our ignorance of both the precise theory giving rise to textures and the precise number of detectable textures in the context of a specific theory. This allows us to compare $\Lambda$CDM and {\em all} models that give rise to textures. Under the assumption of a uniform prior the posterior is simply proportional to the likelihood; results for a different prior on $\bar{N}_s$ could be obtained easily by reweighting the current posterior.

The significance required to favor the $\Lambda$CDM+textures model can be understood by evaluating Eq.~\ref{eq-approx-posterior} using a set of simulated evidence ratios $\rho_{b_i} (\epsilon)$, assuming that two candidate textures have been located in the data. The evidence ratios are chosen to be either low-amplitude and flat in $\epsilon$ (the case where each blob yields no support for the texture model), or Gaussian with varying amplitude (indicating varying degrees of support for the texture model). In all cases, the Gaussians are chosen to peak at the same value, $\epsilon = 5 \times 10^{-5}$, and have the same standard deviation, $\sigma = 5 \times 10^{-6}$. The amplitudes of the Gaussian peaks are selected so that $\int \rho_{b_i}(\epsilon)\,{\rm Pr}(\epsilon)\,{\rm d}\epsilon$ is 1/20, 1, or 20. These values are indicative of weak, intermediate, and strong texture signals, respectively.

\begin{figure}
\begin{center}
\includegraphics[width=8.5cm]{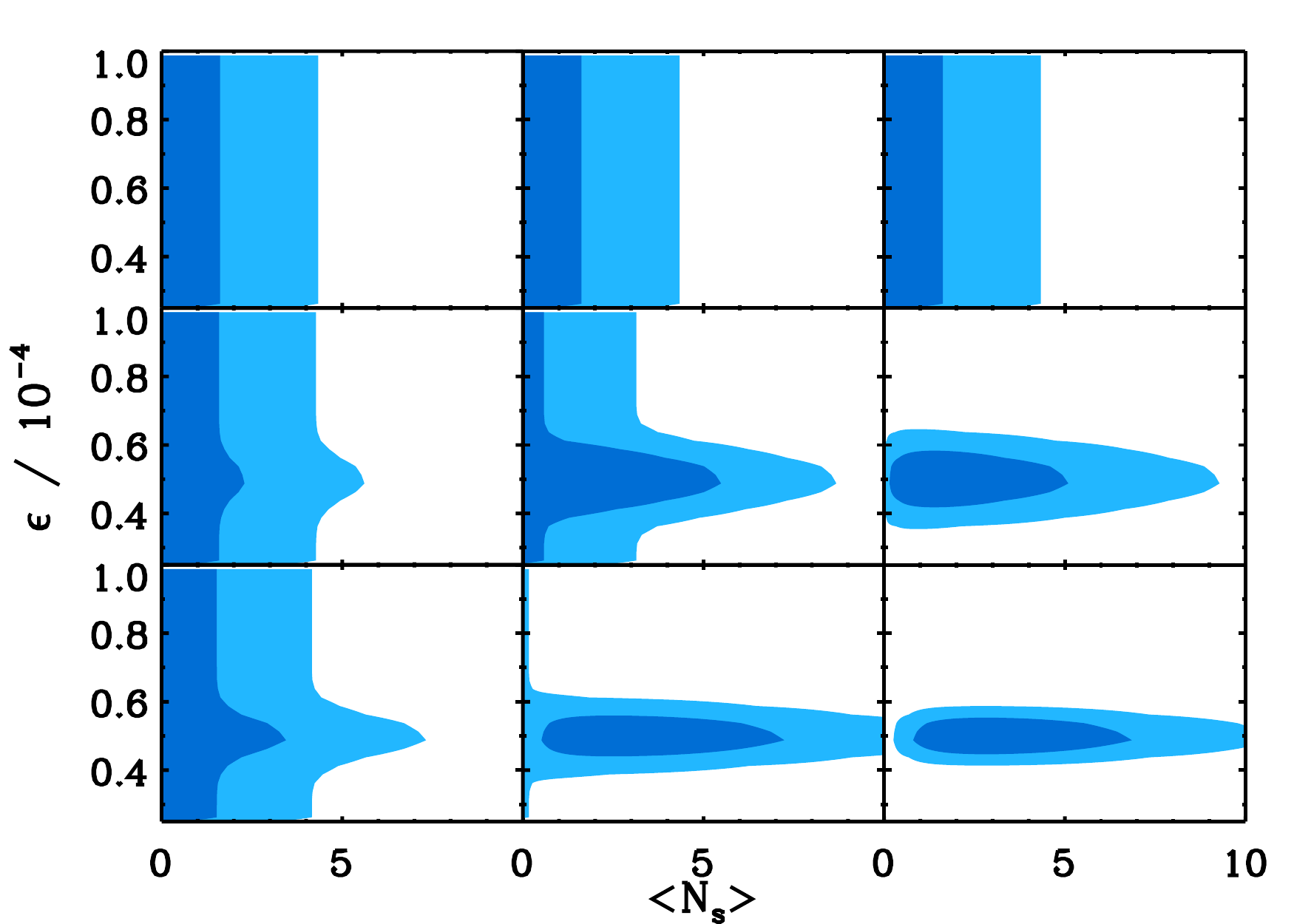}
\caption{Regions containing $68\%$ (dark blue) and $95\%$ (light blue) of the posterior probability distribution, Eq.~\ref{eq-approx-posterior}, for the hypothetical situations described in the text. Each case contains two hypothetical texture candidates. Top row: both candidates are $\Lambda$CDM only, and have low-amplitude, flat evidence ratios, $\rho_{b_i} (\epsilon)$. Middle row: one candidate is ``texture-like'', and has a Gaussian evidence ratio whose amplitude increases from left to right.  Bottom row: both candidates are texture-like; again, their evidence ratio amplitudes increase from left to right. 
  \label{fig-posterior_examples}
}
\end{center}
\end{figure}

The posteriors for all combinations of the simulated evidence ratios are shown in Fig.~\ref{fig-posterior_examples}.  When none of the candidate features support the texture hypothesis (top row), the posterior is exponentially decreasing in $\bar{N}_s$. In this case, we would correctly conclude that pure $\Lambda$CDM is strongly favored, and no constraints on $\epsilon$ could be extracted. When one or two blobs produce a peaked evidence ratio (central and bottom rows), it becomes possible to make a detection. As the amplitudes of the evidence ratios are increased (left to right), the posterior begins to bulge, before ultimately becoming peaked. We would correctly conclude that the data favor $\Lambda$CDM+textures over pure $\Lambda$CDM if a peak in the posterior at $\bar{N}_s \neq 0$ was sufficiently higher than the value of the posterior at $\bar{N}_s=0$. Comparing the central and bottom rows of Fig.~\ref{fig-posterior_examples}, a detection can be made either in the case where there is a single strong candidate, or the case where there is a number of moderately strong candidates (provided each $\rho_{b_i} (\epsilon)$ is peaked in the same range of $\epsilon$). For the one- and two-strong-candidate cases (the centre- and bottom-right plots in Fig.~\ref{fig-posterior_examples}), the peaks of the posterior (after marginalizing over $\epsilon$) are $10$ and $1500$ times that of the value at $\bar{N}_s=0$, respectively.

\begin{figure}
\begin{center}
\includegraphics[width=8.5cm]{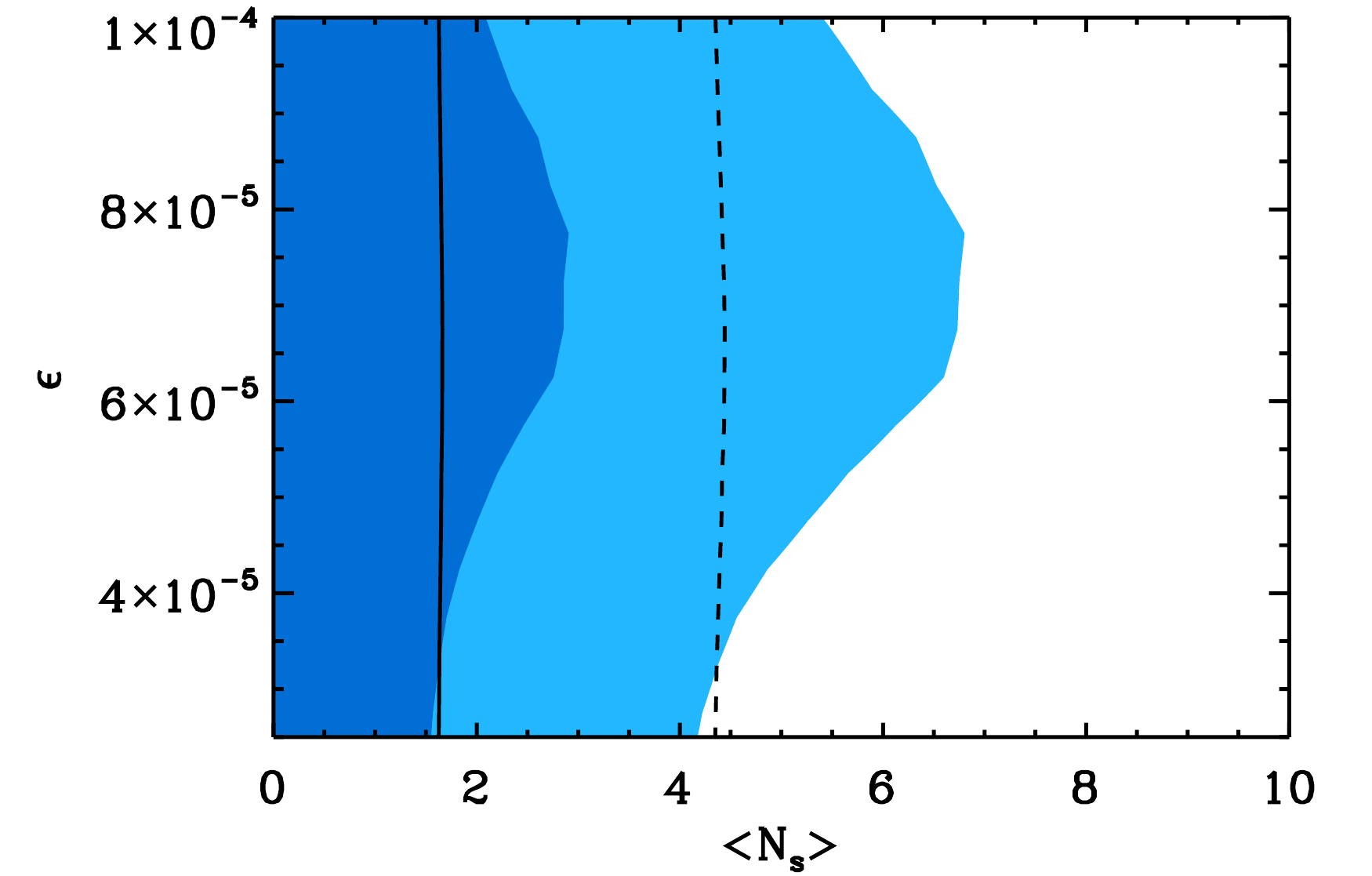}
\caption{Regions containing $68\%$ (dark blue) and $95\%$ (light blue) of the posterior probability distribution, Eq.~\ref{eq-approx-posterior}, for the WMAP7 data, along with the corresponding contours of the $68\%$ (solid line) and $95\%$ (dashed line) of the posterior probability for the end-to-end simulation of the WMAP experiment, based on a pure $\Lambda$CDM model.
  \label{fig-posterior_wmap}
}
\end{center}
\end{figure}

Calculating the posterior for the end-to-end simulation of the WMAP experiment yields the constraints shown as solid and dashed lines in Fig.~\ref{fig-posterior_wmap}. This posterior resembles the top row of Fig.~\ref{fig-posterior_examples}, is peaked at $\bar{N}_s = 0$, and is not significantly different from the input priors on the global texture parameters. We therefore correctly conclude that the end-to-end simulation does not contain textures. 

{\bf Results and conclusions from WMAP.} We perform our analysis on the foreground-subtracted 94 GHz W-band temperature map from the 7-year release of the WMAP experiment~\cite{Larson:2010gs} (prepared by subtracting a model of known astrophysical foregrounds, as described in Ref.~\cite{Hinshaw:2006ia}). The W band has the highest resolution of the five measured by WMAP, with a full-width at half maximum of $0.22^{\circ}$. To minimize the effects of residual foregrounds, we apply the KQ75 mask, which yields a sky coverage of $f_{\rm sky} = 0.706$. The candidate textures are the same as those identified in Ref.~\cite{Feeney:2010dd}, minus one which lies outside our prior on $\theta_c$. The features range in size from $2^{\circ}$ to $17.25^{\circ}$, and we are able to process seven at full WMAP resolution, two at half WMAP resolution and the largest at a quarter WMAP resolution.  Although the lower-resolution computation of the likelihood for these three largest features does result in reduced accuracy, the impact on the overall posterior is minimal as the prior for textures of such large size is very low (cf.\ Eq.~\ref{eq-local-priors}).

Evaluating Eq.~\ref{eq-approx-posterior} yields the posterior for cosmic textures in the WMAP7 data shown in Fig.~\ref{fig-posterior_wmap} as dark- and light-blue regions. The posterior is clearly peaked at $\bar{N}_s = 0$, and we find the marginalized constraint on the expected number of detectable textures to be $\bar{N}_s < 5.9$ (at $95\%$ confidence). We therefore conclude that the WMAP7 data do not warrant augmenting $\Lambda$CDM with textures. The marginalized constraint on the scale of symmetry breaking is found to be $2.6 \times 10^{-5} \le \epsilon \le 1.0 \times 10^{-4}$ (at $95\%$ confidence).

While the posterior is peaked at $\bar{N}_s = 0$, there is also a clear difference between the WMAP7 posterior and that of the end-to-end simulation (over-plotted in Fig.~\ref{fig-posterior_wmap}).  Comparing the WMAP7 posterior to the example plots in Fig.~\ref{fig-posterior_examples}, our result is also consistent with a signal that is present, but too weak to provide a detection. The different shape of the posterior is determined almost entirely by two features, located at $(l = 185^\circ, b = -79^\circ)$ and $(l = 209^\circ, b = -57^\circ)$ in Galactic coordinates, the second of which is the Cold Spot~\cite{Cruz:2004ce,Cruz:2009nd}. As in Ref.~\cite{Feeney:2010dd}, we use information from the multiple frequency bands of the WMAP instrument to confirm that there is no detectable residual foreground contamination in these features. This strongly motivates an analysis with better data, as will soon be provided by the Planck satellite~\cite{2011A&A...536A...1P}, or a better candidate-location technique, such as one utilizing optimal filters~\cite{McEwen:2012uk}. There is also the possibility of including CMB polarization data, as textures would not induce a polarization signal, unlike the primary CMB perturbations~\cite{2011MNRAS.410...33V}). All of these efforts are currently in progress. These and other tests will lead to better constraints on -- or, if a signal is present, a confirmation of -- the texture hypothesis.

\acknowledgements
We are very grateful to Eiichiro Komatsu and the WMAP Science Team for supplying the end-to-end WMAP simulations used in our null tests, as well as Neil Turok for discussions. This work was partially supported by a grant from the Foundational Questions Institute (FQXi) Fund, a donor-advised fund of the Silicon Valley Community Foundation on the basis of proposal FQXi-RFP3-1015 to the Foundational Questions Institute. SMF is supported by the Perren Fund and STFC. Research at Perimeter Institute is supported by the Government of Canada through Industry Canada and by the Province of Ontario through the Ministry of Research and Innovation. HVP is supported by STFC and the Leverhulme Trust. We acknowledge use of the HEALPix package and the Legacy Archive for  Microwave Background Data Analysis (LAMBDA).  Support for LAMBDA is provided by the NASA Office of Space Science.

\bibliography{textures}

\end{document}